\documentstyle[12pt]{article}
\def\oc{\overline c}

\def\utc{\underline{\tilde c}}

\def\uotc{\underline{\tilde{\overline c}}}

\def\uttc{\underline{\tilde{\tilde c}}}

\def\uottc{\underline {\tilde {\tilde{\overline c}}}}

\def\dl{\buildrel\leftarrow\over\delta}
\def\pl{\buildrel\leftarrow\over\partial}
\begin{document}

\title{\LARGE\bf A SUPERSPACE FORMULATION FOR THE BATALIN VILKOVISKY
 FORMALISM WITH EXTENDED BRST INVARIANCE }

\author{ Nelson R.F.
Braga\thanks{\noindent e-mail:braga@if.ufrj.br} and Sergio M. de Souza\\
Instituto de F\'{\i}sica\\
Universidade Federal do Rio de Janeiro\\
RJ 21945-970 - Caixa Postal 68528 - Brasil\\
\date{}}

\maketitle
\abstract
A superspace formulation for the Batalin Vilkovisky formalism (also
called field-antifield quantization ) with  extended BRST invariance
(BRST and anti-BRST invariance ) for gauge theories with closed
algebra is presented.  In contrast to a recent formulation, where
only BRST invariance holds off shell, two collective sets of fields
are introduced and an off shell realization of the extended algebra
in a superspace with two Grassmann coordinates is obtained.  The
example of the Yang Mills theory is also considered.

\vfill
\vspace{1cm}
\newpage
\section{Introduction}

The  Lagrangian BRST quantization procedure of Batalin and
Vilkovisky\cite{BV1,BV2} (BV) is  a very powerful framework for the
quantization of gauge field theories \cite{He,GPS}.
One of it's important features is that when applied to the
 quantization of gauge theories with generators that
form an open algebra it
furnishes a systematic way of building up the ghost structure of the
theory, even for infinitely irreducible theories, where an infinite
chain of ghosts is naturally introduced\cite{BV2}.

Very interesting results have also recently shown up in the
application of the BV procedure to the quantization of anomalous
gauge theories.  Once
one is able to regularize a theory in order to give a well defined
meaning to the quantum Master Equation\cite{TPN} , the anomalies can
be calculated and, by extending the field anti-field
space\cite{GPS,BM}, the Wess Zumino terms can be naturally introduced
in such a way that the theory is BRST invariant at the quantum level.

The standard BV quantization of refs.\cite{BV1,BV2} is based on the
requirement of BRST invariance\cite{BRST} of the total action.  A
Lagrangian quantization procedure, with a larger anti-field
structure, based on the requirement of invariance of the total action
under both BRST and anti-BRST \cite{ABRST} transformations was
presented in \cite{EXT,BVA}. We will call it extended BV formalism
from now on.

A superspace formulation for the BV action at classical level (zero
order in $\hbar$) has recently been given in ref\cite{BD}. In this
article, a BRST invariant formulation was found, by implementing in a
superspace with one grassmann coordinate, an alternative
derivation of the Batalin Vilkovisky action, presented in ref.
\cite{AD}. In  this derivation, the complete set of fields
(fields of the classical action, ghosts, antighosts and auxiliary
fields associated to the original gauge symmetry ) in the theory is
doubled by adding the so called collective fields, increasing in a
trivial way the total symmetry of the action.  The extra symmetries
(shift symmetries) are then fixed in such a way that the BV action is
found, with each of the antighosts of the shift symmetries playing
the role of the associated antifield.

In Ref. \cite{BD} the possibility of implementing a superspace formulation
with extended BRST (BRST and anti-BRST) invariance for the BV action
was also investigated, but using only one set of collective fields, just to
see how far can one go within that framework.  It was found that the
extended BRST invariance would work just on shell.  This result was
in fact expected, taking into account the particular features of
the  extended BV procedure presented in \cite{EXT}. In this
extended formalism there are three different antifields associated to each
field, one generating BRST transformations, one generating anti-BRST
transformations and one generating mixed transformations.  A
derivation of this extended BV action by means of adding trivial
symmetries and then fixing them (collective field approach) was
presented in \cite{EXT2,DJ}.  In this references it is shown that two
collective sets of fields are necessary in order to recover the
complete antifield structure of the extended BV formalism .

The aim of the present article is to present a superspace version for
the extended  BV formalism for
gauge theories with closed algebra at one loop order. As it will
become clear, this will not be a trivial generalization of the
results of \cite{BD} since the gauge fixing structure in this case is
considerably different. In section (2) we will briefly review a
derivation of the BV action with extended BRST invariance by means
of the introduction of two sets of collective fields.  In section (3)
we present our superspace formulation for general gauge theories with
closed algebra. The case of the Yang Mills theory will be presented, as an
example, in section (4). Section (5) is devoted to some concluding
remarks.  Some technical details of the superspace structure for
general theories and also the complete expressions for the superfields
that show up in the Yang Mills example are presented in the appendix.

\section{The BV formalism with extended BRST symmetry}

In this section we will briefly review a derivation of the extended
version of the BV quantization method,  by means of the introduction
of collective fields \cite{EXT2,DJ}. We will use the same notation and
conventions as in these references. Let us consider a Classical
Action ${\cal S}_{class.}(\phi^i)$ of some gauge theory. We introduce
ghosts, anti-ghosts and auxiliary fields associated to the gauge
symmetries of  ${\cal S}_{class.}$, as usual, denoting this enlarged
set of fields as $\phi^A = ( \phi^i, c^\alpha, \overline c^\alpha,
G^\alpha) $. The extended transformations for these fields
are represented in the compact notation:

\begin{equation}
\label{ALG1}
\delta^o_a\,\phi^A = R^{A\, ( \phi )}_a
\end{equation}

\noindent where the index  $a=1,2$ corresponds  respectively to the
BRST and anti-BRST transformations.

Now we introduce two new sets of collective  fields:$\, \varphi^{A1}$
and $ \varphi^{A2}$ and
 substitute the fields $\phi^A$ by $(\,\phi^A - \varphi^{A1}
- \varphi^{A2}\,)$. This enlarges in a trivial way the symmetry content
of the theory, adding to each value of the index $A$ two independent
shift symmetries.
In order to build up a representation for the enlarged BRST algebra
corresponding to the original gauge symmetries plus the new ones,
one associates to the shift symmetries two sets of ghosts: $\pi^{A1}$
and $\phi^{\ast A2}$,
two sets of antighosts  $\phi^{\ast A1}$ and $\pi^{A2}$ and two sets
of auxiliary fields: $B^A$ and $\lambda^A$.  There is a large freedom
in choosing the transformation of the individual fields, since only
$\phi^A - \varphi^{A1} - \varphi^{A2}$ is the relevant quantity. One
possible way of writing out the  extended BRST algebra for this
enlarged system is:

\begin{eqnarray}
\label{ALGEBRA}
\delta_a\,\phi^A &=& \pi_a^A\nonumber\\
\delta_a\varphi^A_b &=& \delta_{ab} \big[ \pi^A_a
-\epsilon_{ac}\phi^{\ast A}_c - R^{A\, (\phi - \varphi_1 - \varphi_2 )\,}_a
\big] + (1-\delta_{ab})\epsilon_{ac} \phi^{\ast A}_c\nonumber\\
\delta_a\,\pi^A_b &=& \epsilon_{ab} B^A\nonumber\\
\delta_a B^A &=& 0\nonumber\\
\delta_a\,\phi^{\ast A}_b &=& -\delta_{ab} \big[
(-1)^a \lambda^A + {1\over 2} (B^A + {\dl R^{A\, (\phi -\varphi_1
-\varphi_2)}_1 \over \delta \phi^B} R^{B\, (\phi^A -\varphi^{A1}
-\varphi^{A2})\,}_2  ) \big] \nonumber\\
\delta_a \, \lambda^A_b &=& 0
\end{eqnarray}

\noindent where the left arrow indicates right derivatives
\cite{DJ}.

The gauge fixing of the shift symmetries is implemented by means of the
gauge fixing action:

\begin{equation}
S_{col.} = - {1\over 4} \epsilon^{ab}\delta_a\delta_b\left[\varphi^{A1}
M^{AB}\varphi^{B1} - \varphi^{A2} M^{AB}\varphi^{B2}\right]
\end{equation}

\noindent where $M^{AB}$ is a c-number matrix defined in such a way
that $\phi^{A} M^{AB}\phi^{B}$ has ghost number zero and even Grassmann
parity.

The gauge fixing of the original symmetry is implemented by
introducing a trivial BRST anti-BRST invariant:

\begin{equation}
S_F = {1\over 2} \epsilon^{ab}\delta_a\delta_b F(\phi^A)
\end{equation}

\noindent where $F$ depends only on the original fields $\phi^A$.

The total action corresponds to

\begin{equation}
\label{C-ACTION}
S = {\cal S}_{class.}[\phi^i] + S_{col.}[ \varphi^{A1},\,\varphi^{A2}]
+ S_F [\phi^A]
\end{equation}

As discussed in detail in \cite{EXT2,DJ}, action (\ref{C-ACTION})
leads to the extended BV formalism, with the (redefined) fields:

\begin{eqnarray}
\phi^{\ast A a\,\prime} &=& (-1)^{\epsilon A} \phi^{\ast B a}
M^{BA} (-1)^{a+1}\nonumber\\
\overline\phi^A &=& {1\over 2} ( \varphi^{A1} - \varphi^{A2} )
M^{BA}
\end{eqnarray}

\noindent playing the role of the three sets of antifields of ref. \cite{EXT}.

\section{Superspace Formulation}

We will now develop a superspace formulation for the extended BV
formalism. As discussed already in \cite{BD}, one of the most
important aspects of this kind of formulation is the way of handling
the antifields.  If we simply try to insert them in some superfields
we would not get the usual BV conditions of relating the antifields
to functional derivatives of some  gauge fixing functional.  In the
collective field approach to the extended BV formalism reviewed in
section ({\bf 2}) there is, in principle, just a standard gauge
theory with no antifields (but with a trivial shift symmetry
structure) .  Only after  a judicious choice of how to gauge fix the
extra shift symmetries is that one recovers the BV formalism,
interpreting some fields as playing the role of antifields.  That is
why we are going to build up a superspace version of section ({\bf
2}).

As explained in the appendix, we consider a
 superspace $(x,\theta^a)$ (with $a =1,2$) and
 build up superfields
in such a way that  the  extended BRST transformations will
be realized as translations in the $\theta^a$ components.
Considering a gauge field
theory involving ,  as in section ({\bf 2}), the set of fields
$\phi^A(x)$ enlarged by the addition of the collective fields
$\varphi^{A1}$ and $\varphi^{A2}$, we introduce the associated
superfields, represented here and in the rest of the article as
underlined quantities: ${\bf \underline \phi}^A (x,\theta^a),
\underline{\bf \varphi}^{A\,1}(x,\theta^a)$ and
$\underline{\bf \varphi}^{A\,2}(x,\theta^a)$ that realize the extended
algebra of eq.(\ref{ALGEBRA}):

\begin{eqnarray}
\label{SUPERFIELDS}
{\bf {\underline \phi}}^A(x,\theta^a) &=& \phi^A(x) + \pi^A_a(x)\theta^a
 +B^A (x) \theta^2\theta^1\nonumber\\
& &\nonumber\\
\underline{\bf \varphi}^{A\,1}(x,\theta^a) &=& \varphi^{A\,1}(x)
+ \left[ \pi^{A1} - \phi^{\ast A2} - R^{A\,(\phi -\varphi^1
-\varphi^2)\,}_1 \right] \theta^1
- \phi^{\ast A1}\theta^2\nonumber\\
&+&\left[ -\lambda^A
+ {1\over 2} (B^A + {\dl R^{A \,(\phi -\varphi^1
-\varphi^2)\,}_1\over
\delta \phi^C} R^{C\, (\phi -\varphi^1
-\varphi^2)\,}_2 )\right] \theta^2\theta^1\nonumber \\
& &\nonumber\\
\underline{\bf \varphi}^{A\,2}(x,\theta^a) &=& \varphi^{A\,2}(x)
+ \phi^{\ast A2}\theta^1 +
 \left[ \pi^{A2} + \phi^{\ast A1} - R^{A\, (\phi -\varphi^1
-\varphi^2)\,}_2\right] \theta^2
\nonumber\\
&+&\left[ \lambda^A + {1\over 2} (B^A + {\dl R^{A\, (\phi -\varphi^1
-\varphi^2)\,}_1\over
\delta \phi^C} R^{C\, (\phi -\varphi^1
-\varphi^2)\,}_2 )\right] \theta^2\theta^1
\end{eqnarray}

The gauge invariance of the Classical action implies  also it's
BRST and anti-BRST invariance. Therefore, in this superspace,
it corresponds just  to a superfield with only the first
component:

\begin{equation}
\label{Classical}
{\cal S}_{class.}\big[ {\underline \phi}^i- \underline{
\varphi^{i\,1}} - \underline{ \varphi^{i\,2}}    \big] = {\cal
S}_{class.} \big[ \phi^i -  \varphi^{i\,2} -  \varphi^{i\,2}\, \big]
\end{equation}

\medskip

The gauge fixing action for the shift symmetries can be written as

\begin{equation}
S_{col.} = \int dx {1\over 2} {\pl \over \partial\theta^2}
{\pl \over \partial\theta^1}
\left[\underline{\bf \varphi}^{A1}
M^{AB}\underline{\bf \varphi}^{B1} - \underline{\bf \varphi}^{A2}
M^{AB}\underline{\bf \varphi}^{B2}\right]
\end{equation}

\medskip
\noindent where $M^{AB}$ is a c-number matrix with the same
properties as in the previous chapter.
Since  $M^{AB}$ is field independent, it is trivially BRST and
anti-BRST invariant, corresponding thus, in our superspace to a
quantity with no $\theta$ components. (${\underline M}^{AB}
= M^{AB} $).

\medskip
The gauge fixing of the original symmetry is obtained using a bosonic
(super) functional  $\underline {\bf F} [{\bf {\underline \phi}}^A
(x,\theta^a )] $ of only the  superfields ${\underline \phi}^A
(x,\theta^a )$.  To arrive at a gauge fixing action that corresponds
to a product of a BRST  and an anti-BRST transformations acting on a
gauge fixing bosonic quantity that depends only on the original
fields $\phi^A (x)$, we must impose a restriction on the functional
$\underline {\bf F}$.  The first component of this functional must
involve just the fields $\phi^A$, that correspond to the first
components of the superfields ${\underline \phi}^A$, otherwise, as we
can see from the component expansion for this superfield (eq.
(\ref{SUPERFIELDS}) it would involve not only the original fields,
but also ghosts, antighosts and auxiliary fields of the shift
symmetry, and would thus not lead to a gauge fixing in the standard
way.  This condition corresponds to the requirement that $\underline
{\bf F}$ does not involve derivatives with respect to $\theta^a$.
Imposing this restriction, the general expansion for the gauge fixing
functional gets:

\begin{eqnarray}
\label{GFA}
\underline {\bf F}[{\bf {\underline \phi}}^A (x,\theta^a )]
&=& F[\phi^A] +\delta_a F \theta^a
+ \delta_1\delta_2 F \theta^2\theta^1\nonumber\\
&=&  F[ \phi^A ] + {\dl F \over \delta \phi^A}
\pi^{Aa} \theta^a \nonumber\\
 &+& \Big[ -{\dl F\over \delta\phi^A} B^A + {1\over 2}
\epsilon^{ab} (-1)^{\epsilon_B +1} [ {\dl \over \delta\phi^A}
{\dl \over \delta\phi^B} F ] \pi^A_a \pi^B_b \Big]
\theta^2\theta^1
\end{eqnarray}

Using this superfunctional we write the gauge fixing of the original
gauge symmetries as

\begin{equation}
S_{F} = \int d^2\theta \int dx \,
\underline {\bf F}\,[{\bf {\underline \phi}}^A (x,\theta^a )\,]
\end{equation}

The total action gets thus

\begin{eqnarray}
\label{ACTION}
 S\,\big[ {\bf{\underline \phi}}^A,\, \underline{\bf\varphi}^{A1},\,
 \underline{\bf\varphi}^{A2}\,\big] &=&
S_{class.}\big[ {\bf{\underline \phi}}^i - \underline\varphi^{i1} -
 \underline\varphi^{i2}\,   \big] +
S_{col.}\big[ \, \underline{\bf\varphi}^{A1},\,
 \underline{\bf\varphi}^{A2}\,\big]
+ S_{F}[\, {\bf{\underline \phi}}^A\,]\nonumber\\
&=& \int d^2\theta\, \,\,\Big\{\,\, S_{class.}\big[ {\bf{\underline \phi}}^i
 - \underline\varphi^{i1} -  \underline\varphi^{i2}\,  \big]
\theta^2\theta^1\nonumber\\
&+&\,\,    {1\over 2} \int dx
\left[\underline{\bf \varphi}^{A1}
M^{AB}\underline{\bf \varphi}^{B1} - \underline{\bf \varphi}^{A2}
M^{AB}\underline{\bf \varphi}^{B2}\right]
+ \int dx \, \underline {\bf F} \big[ {\bf {\underline \phi}}^A \big]
\,\,\Big\}\nonumber\\ & &
\end{eqnarray}

In order to show that this superspace expression would lead to the BV
quantization with extended BRST invariance, as in ref. \cite{EXT},
we can insert the expansions (\ref{SUPERFIELDS}) for the superfields
in (\ref{ACTION}). The result corresponds to equation (\ref{C-ACTION})

\begin{eqnarray}
\label{ACTION2}
 S\big[ {\bf{\underline \phi}}^A,\, \underline{\bf\varphi}^{A1},\,
 \underline{\bf\varphi}^{A2}\,\big] &=&
S_{class.}\big[ {\bf{\underline \phi}}^i  - \underline\varphi^{i1} -
 \underline\varphi^{i2}\,  \big] -
 ( \varphi^{A1} + \varphi^{A2}) M^{AB}\lambda_B\nonumber\\
&+& {1\over 2}  ( \varphi^{A1} - \varphi^{A2}) M^{AB} B_B
+ (-1)^{\epsilon_B + 1} \phi^{\ast A 1} M^{AB} \pi ^{B1}\nonumber\\
&+& (-1)^{\epsilon_B } \phi^{\ast A 2} M^{AB} \pi ^{B2}
+ (-1)^{\epsilon_B } \phi^{\ast A 1} M^{AB}
R^{B\, (\phi -\varphi^1 -\varphi^2)}_1\nonumber\\&+&
 (-1)^{\epsilon_B + 1 } \phi^{\ast A 2} M^{AB}
R^{B\, (\phi -\varphi^1 -\varphi^2 )}_2\nonumber\\
&+& {1\over 2}  ( \varphi^{A1} - \varphi^{A2}) M^{AB}
 {\dl R^B_1\over
\delta \phi^C}^{(\phi -\varphi^1 -\varphi^2)}
 R^{C\, (\phi -\varphi^1 -\varphi^2 )}_2\nonumber\\
&-& {\dl F\over \delta\phi^A} B^A + {1\over 2} \epsilon^{ab}
\pi^{Bb}\big[ {\dl\over \delta\phi^A} {\dl\over \delta\phi^B}
\,F \, \big] \pi^{Aa}
\end{eqnarray}
\bigskip

The vacuum functional is build up functionally integrating over all
the fields involved in (\ref{ACTION2}), represented here as $\mu_\alpha$

\begin{equation}
Z = \int \prod_\alpha [ {\cal D}\mu_\alpha ] exp\{ {i\over\hbar}
S\big[ {\bf{\underline \phi}}^A,\,
\underline{\bf\varphi}^{A1},\, \underline{\bf\varphi}^{A2}\,\big] \}
\end{equation}

Integrating over $\lambda_A$, $\varphi^{A2}$, $B^B$, $\varphi^{A1}$,
$\phi^{\ast A 1}$, $\phi^{\ast A 2}$, $\pi^{B1}$, $\pi^{B2}$, we get
the same result as in reference \cite{EXT}:

\begin{equation}
\label{FP}
Z = \int [ {\cal D}\phi^A]\, exp {i\over\hbar}\{ S_{class.} +
\int dx\,\Big( {\dl F \over    \delta \phi^A}
 {\dl R^A_1\over    \delta \phi^B} R^B_2 +
{1\over 2} R^B_b\big( {\dl \over \delta \phi^A}
{\dl F \over \delta \phi^B}\big) R^A_a \,\Big) \, \}
\end{equation}

\noindent If $F$ depends just on the classical fields $\phi^i$, this
expression will correspond just to the Faddeev Poppov result.

The interpretation of all the steps considered in the present section
for general gauge theories with closed algebra will
be illustrated  in the following section by the
example of the Yang Mills theory.

\bigskip
\section{Example: The Yang Mills Theory}

Let us work out the Yang Mills theory as an illustrative example of
our superspace formulation. The classical Action in this case is

\begin{equation}
{\cal S}_{class.}\,[\,A_\mu\,]\, =\,
 - {1\over 4}\,\int dx \, Tr \left(\, F^{\mu\nu}\, F_{\mu\nu}\, \right)
\end{equation}
\medskip
The set of fields that realize the algebra of extended BRST
transformations corresponds, in this model, to the set of original
gauge fields plus ghost, antighost and auxiliary field: $\phi^A =
(\,A_\mu\,,\, c\,,\, \overline c\,,\, G\, )$.
The standard extended transformations
(without yet the introduction of the collective fields) are:

\begin{eqnarray}
\label{YM}
\delta^o_1 A_\mu &\equiv& R^{\,[ A_\mu]}_1 =  D_\mu c\nonumber\\
\delta^o_1 c &\equiv& R^{\,[ c\,]\,}_1 =  {1\over 2}\,
[\,c,\,c\,]_{_+}\nonumber\\
\delta^o_1\overline c &\equiv& R^{\,[\overline c\,]\,}_1 =  \, G\nonumber\\
\delta^o_1 G &\equiv& R^{\,[ G\,]\,}_1 =  0\\
\,\nonumber\\
\delta^o_2 A_\mu &\equiv& R^{\,[ A_\mu\,]\,}_2 = D_\mu \overline c\nonumber\\
\delta^o_2 c &\equiv& R^{\,[ c\,]\,}_2 = - ( G - [\,c,\overline c\, ]_{_+}
\nonumber\\
\delta^o_2\overline c &\equiv& R^{\,[\overline c\,]\,} =
{1\over 2} [\,\overline c, \overline c\,]_{_+}\nonumber\\
\delta^o_2 G &\equiv& R^{\,[ G\,]\,}_2 =  [\,G, \overline c\, ]
\end{eqnarray}

\medskip
Following the procedure of the previous chapter we enlarge the field
content of the theory, by introducing two collective sets of fields
that will be denoted for simplicity as:
 $\varphi^{A1} =
(\,{\tilde A}_\mu\,,\, \tilde c\,,\,$ $ \tilde{\overline c},\,
 \tilde G\,)$ and
$\varphi^{A2} = (\,\tilde{\tilde A}_\mu\,,\, \tilde{\tilde c}\,,\,$
$\tilde{\tilde{\overline c}}\,,\, \tilde{\tilde G}) $.
The fields $\phi^A$ are replaced by the corresponding $\phi^A -
\varphi^{A1} - \varphi^{A2} $ enlarging the symmetry content of the
theory. In order to represent the enlarged extended BRST algebra,
corresponding to the Yang Mills version of eq.(\ref{ALGEBRA}) we will
need the extra ghosts, antighosts and auxiliary fields that will be
denoted respectively as :

\begin{equation}
\pi^{[ A_\mu\,]\, 1}\,,\,\, \pi^{[ c\,]\, 1}\,,\,\,
\pi^{ [\overline c\,]\, 1}\,,\,\,
\pi^{[ G\,]\, 1}\,\,;\,\,\,\phi^{\ast\,[ A_\mu\,]\, 2}\,,\,\,
\phi^{\ast\,[ c\,]\, 2}\,,\,\,
 \phi^{\ast\,[ \overline c\,]\, 2}\,,\,\,
\phi^{\ast\,[ G\,]\, 2}\,;\,\,\,\nonumber
\end{equation}

\begin{equation}
\phi^{\ast\,[ A_\mu\,]\, 1}\,,\,\,\phi^{\ast\,[ c\,]\, 1}\,,\,\,
 \phi^{\ast\,[ \overline c\,]\, 1},\, \phi^{\ast\,[ F\,]\, 1}\,;\, \,\,
\pi^{[ A_\mu\,]\, 2}\,,\,\, \pi^{[ c\,]\, 2}\,,\,\,
\pi^{[ \overline c\,]\, 2}\,,\,\, \pi^{[ G\,]\, 2}\,;\nonumber
\end{equation}

\begin{equation}
B^{[ A_\mu\,]\, 1}\,,\,\, B^{[ c\,]\, 1}\,,\,\,
B^{ [\overline c\,]\, 1}\,,\,\,
B^{[ G\,]\, 1}\,\,;\,\,\lambda^{[ A_\mu\,]\, 1}\,,\,\, \lambda^{[
c\,]\, 1}\,,\,\,
\lambda^{ [\overline c\,]\, 1}\,,\,\,
\lambda^{[ G\,]\, 1}\,\,.\nonumber
\end{equation}
\bigskip
In order to build up the superfields corresponding to the Yang Mills
version of  eq.(\ref{SUPERFIELDS}), we  must  calculate the crossed
terms that contribute to the last components of the superfields
$\underline{\bf \varphi}^{Aa}$, that means:

\begin{eqnarray}
\label{crossed}
{\dl R^{[ A_\mu]}_1\over \delta \phi^C}\, R^C_2 &=& D_\mu (G - \tilde
G - \tilde{\tilde G} ) - [D_\mu (c-\tilde c -\tilde {\tilde c} ),
\overline c-\tilde {\overline c} -\tilde {\tilde {\overline c}} ]_+
\nonumber\\
{\dl R^{[ c\,]\,}_1\over \delta \phi^C} \, R^C_2 &=& [G - \tilde G
-\tilde {\tilde G} ,c-\tilde c -\tilde {\tilde c}] + [ \overline c -
\tilde{\overline c} -\tilde{\tilde{\overline c}}, (c-\tilde c -\tilde
{\tilde c})^2 ]_+\nonumber\\ {\dl R^{[\overline  c\,]\,}_1\over
\delta \phi^C} \, R^C_2 &=& [G - \tilde G -\tilde {\tilde G} ,
\overline c - \tilde{\overline c} -\tilde{\tilde{\overline c}}]
\nonumber\\
{\dl R^{[ G\,]\,}_1\over \delta \phi^C} \, R^C_2 &=&  0\\ \nonumber
\end{eqnarray}

Inserting these results in (\ref{SUPERFIELDS}) we find the
expressions for the twelve superfields, that will be represented as:
$\underline A_\mu,\,
\underline c\, ,\, \underline {\overline c}\,,\,
\underline G\,;\,\underline { {\tilde A}}_\mu\,,\,
\utc\, ,\,  \uotc\,,\,
\underline {\tilde G}\,;\,
\tilde{\tilde {\underline A }}_\mu\,,
\uttc ,\, \uottc\,,\,
\tilde {\tilde {\underline G}} \,
$.  The complete expressions for these superfields are  given in the
appendix.

\medskip
The superspace version for the classical action is just

\begin{equation}
{\cal S}_{class.}\big[ \underline A_\mu- \underline{
\tilde A}_\mu - \tilde { \tilde {\underline A}}_\mu    \big] =
-{1\over 4} \int dx Tr \big(  F_{\mu\nu}^{\,(\,\underline A_\mu-
\underline{
\tilde A}_\mu - \tilde{ \tilde {\underline A}}_\mu)}
F^{\mu\nu \,(\, \underline A_\mu- \underline{
\tilde A}_\mu - \tilde{ \tilde {\underline A}}_\mu) }\big)
\end{equation}
\medskip

Now let us build the gauge fixing action $S_{col.}$ for  the shift
symmetries. We will choose the only non vanishing terms of $M^{AB}$
to be:

\begin{equation}
M^{[\,A_\mu \,A_\mu \,]} = 1\,\,; \,\, M^{[\,c \,\overline c\, ]} =
1\,\,; \,\, M^{[\,\overline c \,c\, ]} = - 1\,\,;\,\, M^{[\,G\, G\,
]} = 1\,\,;\,\,
\end{equation}

It is easy to see that this matrix satisfies the requirements  of
chapter (2). With this choice, we get:

\begin{equation}
S_{col.} =\int dx {1\over 2} {\pl \over \partial\theta^2} {\pl \over
\partial\theta^1}
\left[ \underline{\tilde  A}_\mu\, \underline{\tilde  A}_\mu\,\,
 +\,\,
\underline{\tilde{\tilde  A}}_\mu \,
\underline{\tilde{\tilde  A}}_\mu
+ 2 \utc \, \uotc\, +\, 2 \uttc\,\, \uottc\, +\underline{ \tilde
G}\,\, {  \underline{\tilde G}} + \underline{\tilde {\tilde G}}\,\,
\underline{\tilde {\tilde G}} \right]
\end{equation}

The gauge fixing action for the original gauge symmetry is of the
same form as eq. (\ref{GFA}). In order to recover the usual Faddeev
Poppov result, we will assume that the gauge fixing functional
$\underline {\bf F}$ depends only on the superfields associated to
the original classical fields $A_\mu$:

\begin{equation}
S_F = \int d^2\theta \int dx {\underline {\bf F}} \big[ \underline
A_\mu \big]
\end{equation}

\noindent and also, as discussed in section ({\bf 3}), that it's
first component involves just $A_\mu$.  If we then consider the
vacuum functional for this theory integrated over all the ghosts,
antighosts and auxiliary fields associated to the shift symmetry and
also over the collective fields , we get the Yang Mills version of
equation \ref{FP}:

\begin{eqnarray}
\label{FPYM}
Z = \int [ d A_\mu]\, [ d \,c\,]\, [ d\,\oc\,]\, [ d G\,]\, exp
{i\over\hbar}\Big\{ S_{class.} +
\int dx\,\Big[ {\dl F \over    \delta A_\mu}
 {\dl R^{[\,A_\mu\,]}_1\over    \delta A_\nu} R^{[\,A_\nu\,]}_2
\nonumber\\
+  {1\over 2} R^{[A_\mu\,]}_b\big[ {\dl \over \delta A_\nu} {\dl F
\over \delta A_\mu}\big] R^{[\,A_\nu\,]}\,\Big]\, \Big\}
\end{eqnarray}

\bigskip
Let us now consider the  gauge fixing functional to be:

\begin{equation}
\overline {\bf F} = -{1\over 2} Tr \underline A_\mu
\underline A^\mu
\end{equation}
\medskip
\noindent with this particular choice, eq. (\ref{FPYM}) gets

\begin{equation}
\label{FPYMP}
Z = \int [ d A_\mu]\, [ d \,c\,]\, [ d\,\oc\,]\, [ d G\,]\, exp
{i\over\hbar}\Big\{ S_{class.} +
\int dx\,Tr \,\Big[ D_\mu G A^{\mu\,a} + \partial_\mu \overline c
(D^\mu c )      \,\Big] \, \Big\}
\end{equation}

\medskip
Corresponding to the usual gauge fixed Faddeev Poppov action for the
Yang Mills field in the Lorentz gauge.

\section{Conclusion}

In the Batalin Vilkovisky Lagrangian quantization procedure one
extends the configuration space by introducing anti-fields
corresponding to all the original fields present in the theory.  In
the case of the extended BV formalism, with extended BRST invariance,
there are three antifields associated to each of the original fields.
In this space, one can define  antibrackets that generate the
extended BRST transformations with the classical BV action as the
generator\cite{EXT}. The complete quantum action is obtained as a
proper solution of the master equation in a power series in $\hbar$
which coincides with the classical BV action (up to renormalization)
when there are no anomalies present. In this paper we have  shown
that this Extended BV quantization, at order zero in loops can be
formulated in a superspace with two Grassmann coordinates.  The
Yang Mills theory provides a nice example of how the gauge fixing
structure works in this superspace formulation of the extended BV.
One realizes, comparing with reference
\cite{BD} the more intricate structure that one needs in the present approach,
with extended BRST invariance.

Another interesting application of these ideas would be to
investigate how higher order corrections (in $\hbar$) associated to
anomalous gauge theories could also be worked out in  this superspace
formulation. This is presently under study.

\vspace{1cm}
\appendix
\renewcommand{\theequation}{A.\arabic{equation}}
\setcounter{equation}{0}
\section*{Appendix}
In this appendix we will work out some details of the structure of
the BRST anti-BRST superspace used in this article. More details can
be found in ref.\cite{SUP}. Our superspace consists of the set
$(x,\theta^a)$ of spatial coordinates $x$ plus two Grassmann
coordinates $\theta^a$, with $a=1,2$. The superfields ${\underline
\phi} (x,\theta^a)$, represented in this article by underlined
symbols, are build up with the following  Taylor expansion in the
Grassmann coordinates:

\begin{equation}
\label{SF}
{\underline \phi} (x,\theta^a) = \phi(x) + \delta_1\phi (x) \theta^1
+
\delta_2\phi (x) \theta^2 +\delta_1\delta_2\phi (x)  \theta^2\theta^1
\end{equation}

\noindent where $\delta_1$ and $\delta_2$ represent BRST and anti-BRST
transformations respectively. In such a way that these
transformations are generated as translations in $\theta^1$ and
$\theta^2$ respectively:

\begin{equation}
\label{TR}
\delta_a \,{\underline \phi}(x,\theta^a) = {\pl \over \partial\theta^a}\,\,
{\underline \phi} (x,\theta^a)
\end{equation}

\noindent where the left arrow indicates that the derivative is acting from the
right.

The transformations $\delta_1$ and $\delta_2$ satisfy the so called
extended BRST algebra:

\begin{equation}
\label{Algebra}
\delta_1^2 = \delta_2^2 = \delta_1 \delta_2 + \delta_2\delta_1 = 0
\end{equation}

\noindent this implies that the last component of a superfield is a
BRST and anti-BRST invariant quantity.

Now considering the example of the Yang Mills theory in section
{\bf 4}, using eqs. (\ref{YM}) and (\ref{crossed})
in the general expression for the superfields (\ref{SUPERFIELDS})
we get:

\begin{eqnarray}
\label{SUPERYANG}
\underline A_\mu &=& A_\mu + \pi^{[A_\mu\,]\, a}\theta^a
 +B^{[A_\mu\,]\,} \theta^2\theta^1\nonumber\\
& &\nonumber\\
\underline { {\tilde A}}_\mu, &=& \tilde A_\mu
+ \left[ \pi^{[A_\mu\,]\, 1} - \phi^{\ast [ A_\mu\,]\, 2} - D_\mu c
\right] \theta^1
- \phi^{\ast [ A_\mu\,]\, 1}\theta^2\nonumber\\
&+&\left[ -\lambda^{[ A_\mu\,]\,} + {1\over 2} (B^{[A_\mu\,]\,} +
D_\mu (G-\tilde G - \tilde{\tilde G}
-[ D_\mu  (c-\tilde c -\tilde {\tilde c} ),
\overline c-\tilde {\overline c} -
\tilde {\tilde {\overline c}} ]_+ \right]
\theta^2\theta^1\nonumber \\
& &\nonumber\\
  \tilde { \tilde {\underline A}}_\mu &=&
\tilde{\tilde {A}}_\mu
+ \phi^{\ast [ A_\mu\,]\, 2}\theta^1 +
 \left[ \pi^{[A_\mu \,]\,2} + \phi^{\ast [ A_\mu\,]\, 1} - D_\mu \overline c
\right] \theta^2
\nonumber\\
&+&\left[ \lambda^{[A_\mu\,]\,} + {1\over 2} (B^{[A_\mu\,]\,} +
D_\mu (G-\tilde G - \tilde{\tilde G}
-[ D_\mu  (c-\tilde c -\tilde {\tilde c} ),
\overline c-\tilde {\overline c} -
\tilde {\tilde {\overline c}} ]_+
\right] \theta^2\theta^1
\nonumber\\
& & \\
\underline c &=& c + \pi^{[c\,]\,}_a\theta^a
 +B^{[c\,]\,} \theta^2\theta^1\nonumber\\
& &\nonumber\\
\underline {\tilde c} &=& \tilde c
+ \left[ \pi^{[c\,]\,1} - \phi^{\ast [c\,]\, 2} + {1\over 2}
[c,c]_{_+}  \right] \theta^1
- \phi^{\ast [c\,]\, 1}\theta^2\nonumber\\
&+&\left[ -\lambda^{[c\,]\,} + {1\over 2} (B^{[c\,]\,} +
 [G - \tilde G -\tilde {\tilde G} ,c-\tilde c -\tilde {\tilde c}]
 + [ \overline c - \tilde{\overline c} -\tilde{\tilde{\overline c}},
(c-\tilde c -\tilde {\tilde c})^2 ]_{_+} )\right]
\theta^2\theta^1\nonumber \\
& &\nonumber\\
\underline {\tilde{\tilde c}} &=& \tilde {\tilde c}
+ \phi^{\ast [c\,]\, 2}\theta^1 +
 \left[ \pi^{[c\,]\, 2} + \phi^{\ast [c\,]\, 1} - (G - [c,\overline c]_{_+} )
\right] \theta^2
+\Big[ \lambda^{[c\,]\,}\nonumber\\
 &+& {1\over 2} (B^{[c]} +  [G - \tilde G -\tilde {\tilde G} ,
c-\tilde c -\tilde {\tilde c}]
 + [ \overline c - \tilde{\overline c} -\tilde{\tilde{\overline c}},
(c-\tilde c -\tilde {\tilde c})^2 ]_{_+}  )\Big] \theta^2\theta^1 \\
& &\nonumber\\
\underline {\overline c} &=& \overline c +
\pi^{[\overline c\,]\,}_a)\theta^a
 +B^{[\overline c\,]\,} \theta^2\theta^1\nonumber\\
& &\nonumber\\
\underline {\tilde {\overline c}} &=& \tilde {\overline c}
+ \left[ \pi^{[\overline c\,]\,1} - \phi^{\ast [\overline c\,]\, 2}
 + G
  \right] \theta^1
- \phi^{\ast [\overline c\,]\, 1}\theta^2\nonumber\\
&+&\left[ -\lambda^{[\overline ,]\,} + {1\over 2} (B^{[\overline c\,]\,} +
 [G - \tilde G -\tilde {\tilde G} ,
 \overline c - \tilde{\overline c} -\tilde{\tilde{\overline c}}]
  )\right] \theta^2\theta^1\nonumber \\
& &\nonumber\\
\underline {\tilde{\tilde {\overline c}}} &=&
\tilde {\tilde {\overline c}}
+ \phi^{\ast [\overline c\,]\, 2}\theta^1 +
 \left[ \pi^{[\overline c\,]\, 2} + \phi^{\ast [\overline c\,]\, 1} -
-{1\over 2} [\overline c,\overline c\,]_{_+} ) \right] \theta^2
\nonumber\\
&+&\left[ \lambda^{[\overline c\,]\,}
 + {1\over 2} (B^{[\overline c\,]\,} +
 [G - \tilde G -\tilde {\tilde G} ,
 \overline c - \tilde{\overline c} -\tilde{\tilde{\overline c}}]
 )\right] \theta^2\theta^1 \\
& & \nonumber\\
\underline G  &=& G + \pi^{[G\,]\,}_a\theta^a
 +B^{[G\,]\,} \theta^2\theta^1\nonumber\\
& &\nonumber\\
\underline {\tilde G} &=& \tilde G
+ \left[ \pi^{[G\,]\,1} - \phi^{\ast [G\,]\, 2} \right] \theta^1
- \phi^{\ast [G\,]\, 1}\theta^2\nonumber\\
&+&\left[ -\lambda^{[\,G\,]\,} + {1\over 2} B^{[\,G\,]\,}  \right]
\theta^2\theta^1\nonumber \\
& &\nonumber\\
\tilde {\tilde{\underline G}} &=& \tilde {\tilde {G}}
+ \phi^{\ast [G\,]\, 2}\theta^1 +
 \left[ \pi^{[G\,]\, 2} + \phi^{\ast [G\,]\, 1} - [G,\overline c] )
\right] \theta^2 \nonumber\\
&+&\left[ \lambda^{[G\,]\,}
 + {1\over 2} B^{[G\,]\,} \right] \theta^2\theta^1
\end{eqnarray}

\noindent {\bf Acknowledgment:} This work is supported in part by
Conselho Nacional de Desenvolvimento Cient\'{\i}fico e Tecnol\'ogico
- CNPq (Brazilian Research Agency).

\end{document}